\documentclass[aps,prb,twocolumn,floats,showpacs,amsmath,amssymb,floatfix,%
superscriptaddress]{revtex4}
\usepackage[dvips]{color}
\usepackage{graphicx}
\usepackage{epsfig}
\usepackage{dcolumn}
\usepackage{bm}
\usepackage{amsmath}
\usepackage{amssymb}
\usepackage{color}

\begin{document}

\title{The dual nature of As-vacancies in 
LaFeAsO-derived 
superconductors:  \\
magnetic moment formation while preserving superconductivity}

\author{Konstantin Kikoin}
\affiliation{School of Physics and Astronomy, Tel-Aviv University
69978 Tel-Aviv, Israel}

\author{Stefan-Ludwig Drechsler}\email{s.l.drechsler@ifw-dresden.de}
\affiliation{Institute for Theoretical Solid State Physics, IFW-Dresden, 
P.O.\ Box 270116, D-01171 Dresden, Germany}

\author{Ji\v{r}i M\'alek }
\affiliation{Institute for Theoretical Solid State Physics, IFW-Dresden, 
P.O.\ Box 270116, D-01171 Dresden, Germany}
\affiliation{Institute of Physics, ASCR, Prague, Czech Republic }

\author{Jeroen van den Brink}
\affiliation{Institute for Theoretical Solid State Physics, IFW-Dresden, 
P.O.\ Box 270116, D-01171 Dresden, Germany}

\date{\today}

\begin{abstract}
As-vacancies (V$_{\rm As}$) in La-1111-systems, which are nominally 
non-magnetic defects, are shown to
  create in their vicinity by symmetry ferromagnetically oriented
  local magnetic moments due to the strong, covalent bonds with neighboring Fe atoms that they break. 
From microscopic theory 
in terms of an appropriately modified 
Anderson-Wolff model, 
we find that the moment formation results in a substantially enhanced 
paramagnetic susceptibility in both the normal and superconducting (SC) state. Despite the
V$_{\rm As}$  act as magnetic scatterers, 
they do 
not deteriorate SC
properties which can even be improved 
by V$_{\rm As}$ by suppressing a 
competing or coexisting commensurate
spin density wave or its remnant fluctuations. Due to the induced
local moments an
$s_{++}$
-scenario is unlikely.
\end{abstract}


\maketitle

Although the general features of the superconducting (SC) and magnetic states 
in the Fe pnictides and chalcogenides are experimentally 
well-documented and theoretical models convincingly outline a predominantly non-phononic 
pairing, 
including multiband effects and unconventional pairing symmetries,  
the detailed mechanism of Cooper pairing in these 
materials still awaits 
elucidation~\cite{Johnston10,PaGree10,AnBoe10}.
From a general point of view, the study
and control of magnetic and non-magnetic defects 
can be helpful in this respect since the way in which they affect the Cooper pairs
depends 
directly on the pairing interactions and symmetries in these multiband systems.
In the pronounced multiband situation with unconventional SC as considered in the Fe-pnictides, 
point-defects/impurities that affect the nonmagnetic interband scattering
 will be detrimental for SC and is in no manner expected to strengthen it.
Note several counterintuitive experimental observations related to the 
presence of As-vacancies (V$_{\rm As}$):
(i) a significantly improved upper critical field slope near a slightly enhanced $T_c$ but
 Pauli-limiting behavior 
above 
30~T and a lacking thereof in "clean" 
samples~\cite{Fuchs08}, (ii) a strongly enhanced spin susceptibility
pointing at
the formation of magnetic moments localized in the vicinity of
V$_{\rm As}$
\cite{Grine11} 
and (iii) a steepened descend of the NMR relaxation rate in the SC state \cite{smart10} as 
compared with "clean" non-deficient As systems.
Here we show that from a theoretical standpoint these observations can be understood in terms 
of a highly unconventional role that
V$_{\rm As}$
play in magnetic and SC 
properties of the As-deficient, optimally doped ${\rm LaFeAs_{1-\delta}O_{1-x}F_x}$. From the 
Anderson-Wolff model
\cite{And61,Wolff61,Midel67}  that we apply,
it turns out that   a nominally non-magnetic 
V$_{\rm As}$
creates a local magnetic moment due its breaking of four strong, covalent Fe-As bonds. This
 moment formation gives rise to the observed strongly enhanced paramagnetic susceptibility in 
both the normal and SC state. At the same time, however, the presence of 
V$_{\rm As}$
does not 
deteriorate
SC
properties and might
even strengthen them.

\begin{figure}[b]
\begin{center}
\includegraphics[width=7.5cm,angle=0]{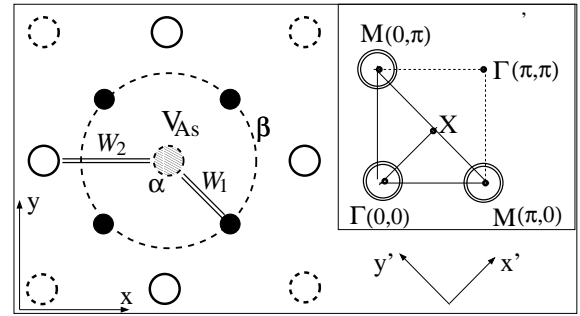}\hspace*{5mm}
\end{center}
\vspace*{-3mm}
  \caption{Bonds
  around a single As vacancy ${\rm V_{As}}$ in a
  Fe-As layer. Fe ions: $\bullet$. As ions above and below the Fe plane:
  solid and dashed $\circ$, respectively. The 
  $dp$-dangling
  bonds 
  (DB)
  between the ${\rm V_{As}}$ 
and its 4 Fe NN
  neighbors forming the orbital $\beta$ are labeled as $W_1$; 
$pp$-DB
between 
${\rm V_{As}}$ orbital $\alpha$ and its 4
 neighbors in As tetrahedra are labeled as $W_2$. Inset: fragment of the BZ for 
a single
FeAs layer (dashed lines) and the folded BZ for the Fe sublattice (full lines).
The axes for full lattice and Fe sublattice are denoted as (x,y) and (x',y'),
respectively.}  \label{Fig_1}
\end{figure}

\begin{figure}[b]
\begin{center}
  \includegraphics[width=7.5cm,angle=0]{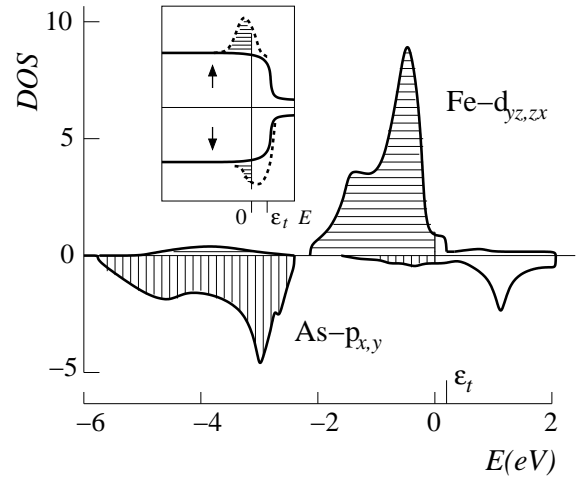}\hspace*{5mm}
\end{center}
\vspace*{-3mm}
  \caption{Partial DOS for ${\rm Fe}_{xz,yz}$ and ${\rm As}_{x,y}$ states
(following the DFT-TB approximation
\cite{EsKo,YYO}). $E=0$ corresponds to $\varepsilon_F$. Inset: vacancy induced peaks in the DOS near
$\varepsilon^{}_t$ (the top of the hole $d$, see also Fig.\ \ref{fig_3})} \label{fig_22}
\end{figure}

An V$_{\rm As}$ 
in the anisotropic $\rm FeLaAs_{1-x}OF$ system can be considered as 
a 
dangling-bond 
(DB)
defect in its Fe-As triple
layers, which 
electronically
can be viewed as 
single layers
(see Fig.\ \ref{Fig_1}).
The ${\rm V_{As}}$, treated as a missing ${\rm As}^{3-}$, generates 
DB
with
4 neighboring Fe ions in the central square plane and 4 neighboring As ions 
in
tetrahedral coordination. In a minimal tight binding model these bonds are formed by
the $p_{x,y}$ and $d_{xz,yz}$ orbitals. The corresponding
$dp$ and $pp$ hopping integrals are denoted as $W_1$ and $W_2$, respectively.
This model reflects the basic features of the band structure of LaFeAsO
(see, e.g., \cite{Anis09,EsKo,YYO}). According to density functional theory
combined with a tight binding analysis (DFT-TB), the Fe-related
$3d$ bands range 
from
 -2 eV to +2 eV
around the Fermi level $\varepsilon_F=0$. The bonding and antibonding $4p_{x,y}$ states contribute to
As-related bands deep below $\varepsilon_F$ and the nearly empty bands above it.
The partial density of states (DOS) for these states is shown 
schematically in Fig.\ \ref{fig_22}.
The empty states in the 
very 
vicinity of $\varepsilon_F$ correspond to 
hole ($h$) 
and electron ($el$)
pockets, and the pronounced step in the DOS at the upper band edges arises 
due to
the nearly 2D character of these $dp$-bands.

In view of this structure of the energy spectrum, the experimentally reported
influence of ${\rm V_{As}}$ defects on the magnetic properties of doped LaFeAsO looks
especially puzzling. Indeed, the overlap between the $p_{x,y}$ states and
$d_{yz,zx}$ state is noticeable only near $\varepsilon_F$. The mechanism, which may be responsible
for the formation of local magnetic moments in defect cells is the
influence of broken $dp$ valence bonds described by the transfer integral $W_1$
(Fig.\ \ref{Fig_1}) on the states near $\varepsilon_F$. We conclude from the geometry
of the DB
and the structure of the band spectrum that the
${\rm V_{As}}-$related defect potential disrupts mainly
the hybridization with the $d_{yz,xz}$ partial component of the
$d$ states and note that this transfer integral
should  mainly contribute to the intraband scattering within the $h$-pocket
of the  Fermi surface (FS), which is formed mainly by the $d_{yz,xz}$ states.

The resulting Hamiltonian of this minimal model is
\begin{eqnarray}\label{h0}
 H &=& H_d + H_p + H_{dp} + H_{vd} =  \sum_{{\bf k}\sigma} \varepsilon_{k,d} d^\dag_{{\bf k}\sigma}d^{}_{{\bf k}\sigma} \nonumber\\
  &+&   \sum_{{\bf k}\sigma} \varepsilon_{{\bf k},p} p^\dag_{{\bf k}\sigma}p^{}_{{\bf k}\sigma}
+ \sum_{{\bf k}\sigma} \left( V_{dp}({\bf k}) d^\dag_{{\bf k}\sigma}p^{}_{{\bf k}\sigma} + {\rm H.c.}\right)\nonumber\\
&+& W_{1}\sum _{j\sigma} \left( d^\dag_{j\sigma}p^{}_{0\sigma} + {\rm H.c.}\right).
\end{eqnarray}
The first three
terms describe the nearly filled $d_{xz,yz}$ band
$\varepsilon_{k,d}$, the nearly empty $p_{x,y}$ band $\varepsilon_{k,p}$\textcolor{red}{,}
and the hybridization
between them, respectively. $H_{vd}$
is the V$_{\rm As}$-induced
hybridization between the Wannier $p$-state in the defect cell
labeled as "0" and the $d$-band. An orbital splitting will be irrelevant 
in what follows and we do not account for it here. It
results in a double degeneracy of each band.
The defect potential is given by the $dp$-DB
characterized by the coupling constant 
$W_{\beta\alpha}=\langle\beta |W_1|\alpha\rangle$ (see Fig.\ \ref{Fig_1}).

The energy spectrum of defect related states is determined by the secular equation
(see also Ref.\ \onlinecite{epaps})
\begin{equation}\label{sec}
1 - \sum_\alpha W_{\beta\alpha}G^0_{\alpha\alpha}(\omega)W_{\alpha\beta} G^0_{\beta\beta}(\omega) =0 ,
\end{equation}
 The local Green's functions (LGF) are defined as
$
G^0_{\alpha\alpha}(\omega) =  \sum_{\bf k}\frac{\langle \alpha |{\bf k}p\rangle 
\langle {\bf k}p| \alpha\rangle} {\omega - \varepsilon_{{\bf k}p}}$
and
$
G^0_{\beta\beta}(\omega) =  \sum_{\bf k}\frac{\langle \beta |{\bf k}d\rangle 
\langle {\bf k}d|\beta\rangle} {\omega - \varepsilon_{{\bf k}d}}.
$
The $T$-matrix for the scattering in the $d$-band is
$
 {\cal T}_{\bf kk'}= \frac{F_\beta({\bf k})W_\beta(\omega)F_\beta({\bf k'})}
{1-W_\beta(\omega)G^0_{\beta\beta}(\omega)},
$
with the structure factor $F_\beta({\bf k})=\langle {\bf k}d|\beta\rangle$ and 
$W_\beta=\sum_\alpha W_{\beta\alpha}G^0_{\alpha\alpha}(\omega)W_{\alpha\beta} $.

\begin{figure}[b]
\begin{center}
   \includegraphics[width=6.0cm,angle=0]{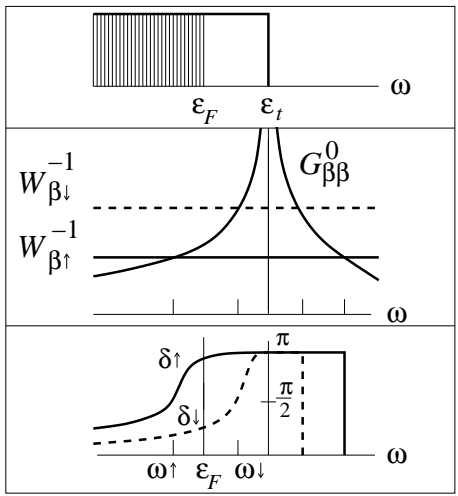}\hspace*{5mm}
\end{center}
\vspace*{-3mm}
\caption{Upper: DOS near $\varepsilon_t$. Middle:
Graphical solution of Eq. $U^{-1}_{\alpha\sigma}(\omega) ={\rm Re} G^0_{\alpha\alpha}(\omega)$
for spin $\uparrow$ (solid line) and spin $\downarrow$ (dashed line). Lower:
Frequency dependent phase shifts $\delta_\uparrow(\omega)$ and
$\delta_\downarrow(\omega)$ (solid and dashed curves, respectively). The energies $\omega_\sigma$
mark the positions of resonances $\delta_\sigma=\pi/2$.
}
\label{fig_3}
\end{figure}

From
general properties of the LGF \cite{Eco},  we conclude that
for repulsive vacancy potential $W_1>0$ the effective potential $U_\alpha(\omega)$ is also
positive for $\omega$ close to the top of the
band $\varepsilon_{a}$.  Then we anticipate strong intraband
scattering in the $h$-pocket of the FS
due to the nearly 2D electronic structure. 
In  this case the DOS at the top $\varepsilon_t$
of the $h$-band is constant $\nu_0$, and this step-like singularity results in the 
logarithmic divergence of the LGF.
\begin{equation}\label{edgesing}
{\rm Re }G^0_{\beta\beta}(\omega \to \varepsilon_t) \propto
 \nu_0\ln \left[ |\omega - \varepsilon_t|/D \right] 
 \quad .
\end{equation}
Here $D$ is an
effective bandwidth
(see e.g.\ Ref.\ \onlinecite{Eco}). Such an edge-singularity of the LGF
means that the resonance
(the zero in the denominator of 
the $T$-matrix should appear
near
$\varepsilon_t$,
even if the scattering potential is weak (Fig.\ \ref{fig_3}, middle panel) so that in any case the
impurity scattering in the $h$-pocket is close to the unitarity limit: the scattering
phase $\delta(\varepsilon_F)$ is close to $\pi/2$. Thus, the DB
induce a scattering
in the $(xz,yz)$ channel, which generates mainly {\it intraband} scattering
in the $h$-pocket.

Now
we turn to the magnetic structure of a
$[{\rm V_{As},Fe_4}]$ defect. The V$_{\rm As}$
itself is not
magnetically active,
but involving
its four Fe NN and possibly also four NNN
in the complex defect shown in Fig.\ \ref{Fig_1} changes the situation dramatically.
Indeed, the necessary precondition for the formation of localized magnetic states in
a system with itinerant electrons is the presence of a noticeable short-range spin-dependent
interaction which may overcome the kinetic energy of the electrons. Fe ions bound
with V$_{\rm As}$
may be the source of such interactions due to the on-site Hubbard repulsion
$U_0$ in their 3$d$ shells. To model
this effect, one has to add the 
term
$U n^{}_{\beta\uparrow}n^{}_{\beta\downarrow}$ to the
Hamiltonian (\ref{h0}).
Here $U$ is the intracell Coulomb repulsion integral for the
''molecular orbital`` $|\beta\rangle$ (see Fig.\ \ref{Fig_1}).
We expect
$U_b \lesssim U \lesssim U_0$ due to the slight delocalization of the  $3d$ wave functions
because of  $dp$-mixing  and a reduced screening due to the missing V$_{\rm As}$
\cite{Anis09}, 
where $U_b$ denotes the Coulomb
repulsion on bulk Fe-sites. Treating
this interaction in a mean-field manner results in 
an additional spin-dependent term in the local scattering potential~\cite{foot2} so that
$W_\beta \to W_{\beta\sigma}= W_\beta+ U\bar n_{\beta,-\sigma}$.

One may expect that
the broken $dp$-valence bonds in the
presence of a
short range Coulomb repulsion in the $d$-shells of Fe ions involved in the
formation of the defect can
result in a {\it spin-dependent} scattering similarly to the well known  Wolff impurity model,
\cite{Wolff61,Midel67} which explained the appearance of localized moments in
metals with potential scatterers in the same way as the Anderson impurity model \cite{And61}
explained this effect in metals with resonant scatterers. A localized
moment in a system of itinerant electrons arises when the self-consistent solution with
$\bar n^{}_{\beta\uparrow}\neq \bar n^{}_{\beta\downarrow}$ of the Dyson equation for
the electron GF
exists, provided the repulsive potential exceeds some critical value
(see also Ref.\ \onlinecite{epaps}).
This means that two defect-related narrow
peaks arise in the DOS
near
$\varepsilon_F$
(see the inset in Fig.\ \ref{fig_22} and the lower panel in Fig.\ \ref{fig_3}).
 In terms of the scattering phase shifts
displayed in the lower panel of Fig.\ \ref{fig_3} the local magnetic order induced by the defect
$[{\rm V_{As},Fe_4}]$ means that $\delta_\uparrow(\varepsilon_F) >\delta_\downarrow(\varepsilon_F)$.
 Both phases $\delta_\sigma \lesssim\pi/2$, which means that the magnetic scattering is strong and not
too far from the unitarity limit.

Thus, we have found that the nominally nonmagnetic ${\rm V_{As}}$ defect can give rise to
the appearance of localized moments formed by
states in the $h$-pocket due to the quasi-2D
character of the electronic band spectrum in ferropnictides. This
explains
the observed strong enhancement of the magnetic susceptibility $\chi(0)$  in an
As-deficient La-1111 system~\cite{Grine11}, with an
enhancement factor~\cite{Midel67}
of
$
\chi / \chi_p -1 \approx cU\chi_p /\left( 1-U \chi_l \right),
$
where $\chi_p$ is the Pauli-spin
susceptibility of the pristine La-1111 compound, $c$ denotes the
$\rm V_{As}$ concentration, $\chi_l =\langle S_\beta,S_\beta\rangle$ is the local susceptibility
at the defect site.

While the magnetic moment formation around V$_{\rm As}$
sites explains the substantially enhanced paramagnetic susceptibility, 
one might expect that with respect to SC,
it opens up a
Pandora's box. Why is the magnetic moment formation in the $h$-pockets not detrimental for
SC~?
The most straightforward explanation should be related to cases
when a competing or coexisting commensurate stripe-like 
spin density wave (CS-SDW) phase or its short-range fluctuations are
still present which detrimental effect on SC is well-known.
Then, the strongly enhanced
scattering of intinerant electrons from the $h$-pockets
in their intraband channel 
found above, 
will help to 
suppress further
its influence and enhance $T_c$ \cite{Vavilov11}.
This effect becomes weaker in the strongly overdoped
region where the SDW suppression by the doping itself is more and more pronounced,
Obviously, also the appearence of relatively large magnetic 
defects with ferromagnetically ordered large local moments provides
a strong perturbation for such a CS-SDW found in undoped
clean samples.
As a result at low $V_{\rm As}$ concentrations
its
transformation  
to
an inhomogeneous magnetic state with discommensurations, 
or to 
a
spin-glass type phase \cite{Grinenko2}
are expected. For  them it is much easier
 to establish  a 
coexistence of magnetism and SC.
 The next intriguing puzzle is then why is the formation of a magnetic moment 
 in the $h$-pockets 
not detrimental for the SC as in usual 
 $s_{++}$-SC? 
The possibility for a qualitatively new solution to this paradox
relies on the observation that
neither the standard theory of doped single-band SC, nor the available approaches to imperfect multiband SC
\cite{Preo96,Parker08,NgAv09,Zhang09,LiWang09,Akbar10,KarOg10,OkaMi11,JiGuLi11,Efremov11}
can be used to address this issue in Fe-pnictide superconductors.
\begin{figure}[t]
\begin{center}
  \includegraphics[width=7.5cm,angle=0]{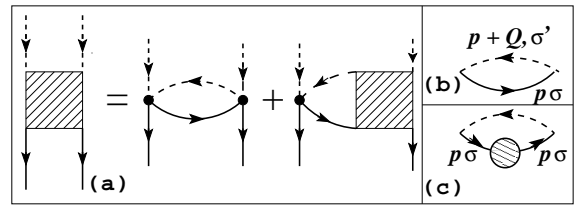}\hspace*{5mm}
\end{center}
\vspace*{-4mm}
  \caption{Diagrammatic representation for the anomalous vertex $\Gamma$ (a), the susceptibility
 $\chi_{\rm sdw}$ responsible for the SDW fluctuations (b) and defect related corrections to
$\chi_{\rm sdw}$ (c). Dashed and solid lines correspond to the $el$- and $h$-propagators,
respectively. 
{\Large $\bullet $ }
are the anomalous Coulomb vertices $u_3$ (in notations of
Ref.\ \onlinecite{MaChub10}). Shaded circle stands for the $T$-matrix
 (see the text for a more detailed explanation).} \label{fig_4}
\end{figure}
Since the
mechanism of SC
in Fe-pnictides is not
established as yet, we start 
with
general remarks on the role of ${\rm V_{As}}$ related defects in 
our SC 
for which actually also the symmetry of the order parameter is under debate, which
may in principle be different in
different 
Fe-based materials and even depend on the doping
type~\cite{Johnston10,PaGree10,Chubuk11,Flet09,Hanaguri10,Hashimoto10}.
To be definite, we will consider
here only the most with a nodeless order parameter $\Delta_{sc}$ having 
opposite signs for the
$h$- and the $el$-pockets~\cite{MSJD08,Kuroku08}.
In accordance with the theory of $s_{\pm}$ 
multiband SC in
pnictides,~\cite{Chubuk11,MaChub10}
the pairing
is mainly
given by the anomalous vertices shown in
Fig.\ \ref{fig_4}
containing the interband spin susceptibility $\chi_{\rm SDW}(\bf q,\omega)$ (Fig.\ \ref{fig_4}b)
as a main element.  The spin density wave (SDW) fluctuations with the vector $\bf q = p+Q$
close to the Umklapp vector connecting
$\Gamma$ and $M$ in the BZ (Fig. \ref{Fig_1}) mediate the Cooper pairing even
in the absence (or suppression) of attractive interactions within $el$- and $h$-pockets of
the BZ. This means that the main
contribution of
the magnetic scattering to the $h$-propagator comes in via
$\chi_{\rm sdw}$ (Fig.\ 4c), represented by the intraband $T$-matrix ${\cal T}_{\bf pp}$.
It is seen from 
this diagram that we deal with magnetic scattering
{\it without} spin flips,
 which  creates narrow local resonance levels below and above 
 $\varepsilon_F$ in the $h$-pockets
and modifies the $h$-propagators in the bubble $\chi_{\rm sdw}({\bf q},\omega)$.

Significantly SDW-affected Cooper pairing may be realized due to the
almost singular behavior of
 $\chi_{\rm sdw}({\bf Q},\omega)\propto \nu_0\ln[D/max\{\omega,\epsilon,\gamma\}]$, where $D$ is the energy
interval where the nesting conditions are approximatily satisfied, $\epsilon$ and $\gamma$ are
parameters characterizing the imperfection of nesting 
(including modifications of the magnetic response due to the
presence of the $V_{\rm As}$) 
and the electron damping due to imperfection of
a real
crystal. The contribution of V$_{\rm As}$ defects to $\epsilon$ are
$\propto \rm Re {\cal T}(\omega)$. Besides, As vacancies can act as 
dopants.\cite{foot3} The net result
of changed $\epsilon$ is
not known {\it a priori}: based on the interplay between these physical effects
the nesting conditions may either slightly improve or worsen, or just 
remain unaltered. The damping $\gamma \lesssim c|\varepsilon_F-\omega_\sigma|$ is
efficient only provided $\gamma > \epsilon$. 
The main point is that either of
these mechanisms cannot radically
reduce the SC
$T_c$. Besides, magnetic resonances give their own contribution
$\delta\chi_{\rm sdw}({\bf q},\omega)$ not related to the nesting. Transitions from the local states
in the 
{\it h-}pocket to the empty states in the 
{\it el-}pocket result in
$\delta\chi_{\rm sdw}({\bf q},\omega)\propto c\nu_0\ln[D/(\omega-|\omega_\sigma|)]$. This contribution
favors 
$s_{\pm}$-pairing. To summarize,
from an analysis of
the consequences of V$_{\rm As}$ defects and their magnetic moment formation
on the SC
Fe-pnictides, it follows that their presence can even be
more constructive than detrimental for 
$s_\pm$ SC
that is mediated by SDW fluctuations.
In this context it is interesting to note the related 
conclusions on the role of intraband magnetic scattering
\cite{LiWang09}.
The 
arguments given above are also
in the spirit of the "Swiss cheese" model \cite{Parker08}, which presumes that 
the 
defect related
bound states contribute to the subgap DOS and do not suppress $T_c$ completely.

The spin susceptibility $\chi_{\bf q}(\omega)$, is responsible also
for the $1/T_1$ NMR spin-lattice relaxation rate, which changes from a
nearly power-law dependence $\sim T^3$
in As stoichiometric samples to more steep and close to an exponential one in As-deficient samples.
The analysis above suggests that
these changes relate
to the interplay between the mid-gap
states stemming from
non-magnetic scattering induced by F doping and those inserted by magnetic defects
V$_{\rm As}$ (cf.\ Ref. \onlinecite{Zhang09}).
The interplay
of magnetic vortices with both kinds of impurities, will affect the $1/T_1$ rate
\cite{JiGuLi11,Zhou11}. A detailed analysis
will be published elsewhere.


We have shown, in conclusion, that As-vacancies form highly nontrivial defects
in 1111 
Fe-pnictide superconductors 
which strongly modify their
physical properties in both the normal and the SC
states. Being nominally non-magnetic in nature, they are nevertheless responsible
for the formation of relatively large local magnetic moments on the Fe-sites
 surrounding the vacancies, which give rise to enhanced spin susceptibility in the normal state
and Pauli-limiting behavior in the  SC
state.\cite{Grine11} The behavior
at low fields is unusual too, due to
scattering properties remarkably different
from those of usual magnetic impurities in standard single band and dirty
multiband $s_{\pm}$ SC.
Controling these defects 
can be
helpful
to improve
the understanding of
real
$s_{\pm}$ systems,
the electronic
structure, and correlation effects in the pnictides, in general.
In particular, for 122-pnictides
with As vacancies
\cite{Ni08} and Fe-chalcogenides
one expects a
similar but somewhat weaker scattering
effect due to
the larger electronic dispersion perpendicular to the FeAs planes.
An 
analysis on various
types of
point-defects will also be
of considerable
interest, since
our results imply that
they may strongly affect the physical properties
of real pnictide materials
even at small defect
concentrations \cite{Grinenko2}.

We thank the DFG SPP 1458 and the Pakt f\"ur For-
schung (IFW-Dresden)
 for support and appreciate 
discussions with
H.\ Eschrig$^\dagger$, M.\ Kiselev, K.\ Koepernik,
D.\ Efremov, 
R.\ Kuzian,  
G.\ Fuchs, 
and J.\ Engelmann

\newpage
\widetext
\vspace{1.5cm}
\hspace{5cm}
{\bf \large APS Supplementary Online Material:}\\
\centerline{\bf \large "
The dual nature of As-vacancies in LaFeAsO-derived superconductors:}
\centerline{\bf \large
magnetic moment formation while preserving superconductivity"}
\vspace{0.5cm}

\centerline{Konstatin Kikoin$^1$, Stefan-Ludwig Drechsler$^2$, Ji{\v r}i M\'alek$^{2,3}$, and Jeroen van den Brink$^2$}

\vspace{0.3cm}
\noindent
$^1$ School of Physics and Astronomy, Tel-Aviv University,
69978 Tel-Aviv, Israel\\
$^2$ IFW-Dresden, P.O.\ Box 270116, D-01171 Dresden, Germany\\
$^3$ Institute of Physics, ASCR, Prague, Czech Republic \\


\setcounter{equation}{0}
\setcounter{figure}{0}
\setcounter{table}{0}

\renewcommand{\theequation}{S\arabic{equation}}
\renewcommand{\thefigure}{S\arabic{figure}}
\renewcommand{\thetable}{S\arabic{table}}
\setcounter{equation}{0}
\setcounter{figure}{0}
\setcounter{table}{0}

\unitlength1.0cm
\noindent
In the present Supplementary part we provide the reader with
the details of the calculations of the
[V$_{\rm As}$,Fe$_4$]-defect related Green's function and the spectrum of local electron excitations.  
We present also an outlook about the interaction of
As-vacancies with an SDW or its fluctuations in a nonmagnetic state.\\

\vspace{0.5cm}

\centerline{\bf Derivation of the secular equation for the defect related states}

\vspace{0.3cm} 

Here we calculate the local Green's  function of electrons in the Fe 
3$d$ band containing a defect from 
four broken valence bonds with the $4p$ bands related 
to the As sublattice.
In the Bloch representation the defect 
related scattering Hamiltonian $H_{vd}$
reads as
\begin{equation}\label{h1}
 H_{vd}= W_1 \sum_{{\bf k}\sigma}\left[ F({\bf k})d^\dag_{{\bf k}\sigma} p^{}_{0\sigma}  
+ {\rm H.c.}\right] ,
\end{equation}
where $F({\bf k})$ is the structure factor depending on the symmetry of the defect potential 
(see below). The scattering term $H_{vd}$ describes the perturbation inserted by  
a ${\rm V_{As}}$ in the $d$-band, and our task is to calculate the reconstruction of the states 
in the hole pocket induced by this perturbation. 

The Green's  functions for the band Hamiltonian form the matrix ${\sf G}(\omega)$:
\begin{equation}\label{matg}
{\sf G}(\omega) = \left( 
\begin{array}{cc}
G_{dd}(\omega) & 0\\
0 & G_{pp}(\omega)
\end{array}
\right)
\end{equation}
(here the spin index is temporarily omitted). 
The perturbation $\sim W_1$ is nonzero in a limited space around an 
As-vacancy. The states within this 
cluster are described in a local basis $(\alpha,\beta)$ formed by projecting the states $(p,d)$ 
onto the perturbed region. Then the perturbation is described by the matrix $\sf W$,
\begin{equation}\label{matw}
{\sf W}=
\left(
\begin{array}{cc}
0 & W_{\beta\alpha}\\
W_{\alpha\beta} & 0
\end{array}
\right)
\end{equation}
so that only the $dp$ bonds are broken in the defect cluster. 

The local basis  obeys the point symmetry of the 2D lattice. In the simplest
tight-binding approximation the states $|\alpha\rangle$ are the orbital states $p_x,p_y$ 
centered at the As-vacancy site ``0''. 
Then the states $|\beta\rangle$ are the ``molecular'' orbitals formed by 
$d_{yz,zx}$ orbitals centered at the sites 1,2,3,4 surrounding the As-vacancy site ``0'' and transforming
along the same irreducible representation of the point group as the states $|\alpha\rangle$, namely
the combinations 
$$|\beta\rangle = |d_1\rangle -|d_2\rangle +|d_3\rangle -|d_4\rangle,$$
where the Fe-sites in the first coordination sphere around the vacancy are enumerated as 1,2,3,4. 
Next we construct the
secular matrix ${\sf Q}(\omega) = {\sf 1} - {\sf WG}(\omega)$ and project this matrix on the
local basis $\{\alpha, \beta \}$, ${\sf Q} \to \widetilde {\sf Q} $.
\begin{equation}\label{tmatrix}
\widetilde {\sf Q}(\omega) =
\left(
\begin{array}{cc}
 1 & -W_{\alpha\beta}G^{0}_{\beta\beta}(\omega) \\
-W_{\beta\alpha}G^0_{\alpha\alpha}(\omega) & 1
\end{array} 
\right) \quad .
\end{equation}
We derive from (\ref{tmatrix}) the secular equation ${\rm det}\, \widetilde {\sf Q}(\omega) = 0$.
Since we are interested in the scattering in the band $\beta$ related to the
$d$-states, we project this secular equation on the subset  $\langle \beta|\ldots |\beta\rangle$. 
\begin{equation}\label{sec}
1 - \sum_\alpha W_{\beta\alpha}G^0_{\alpha\alpha}(\omega)W_{\alpha\beta} G^0_{\beta\beta}(\omega) =0 .
\end{equation}
$W_{\beta\alpha}=\langle\beta |W_1|\alpha\rangle$. 
Comparing Eq.\ (\ref{sec})  with a Slater-Koster-like equation $1-WG^0=0$ for a single band
defect characterized by the local potential $W$
we note that in our problem 
this potential is substituted for the non-local potential $W_\beta$:
\begin{eqnarray}\label{spot}
&& 1 - W_{\beta}(\omega)G_{\beta\beta}(\omega) =0 \\
&& W_\beta (\omega) =\sum_\alpha W_{\beta\alpha}G^0_{\alpha\alpha}(\omega)W_{\alpha\beta} \ . \nonumber
\end{eqnarray}
The solution of Eq.\ (\ref{sec}) provides us with the information about 
the scattering phase $\delta_\beta(\omega)$
\begin{equation}\label{phshift}
 \tan \delta_\beta(\omega) = - \frac{{\rm Im \,det}\,\widetilde Q(\omega)}
{{\rm Re \,det}\,\widetilde Q(\omega)} \ .
\end{equation}
 which characterizes the strength of the
defect potential at the Fermi level $\omega=\varepsilon_F$.
The $T$-matrix for the scattering in the $d$-band is  
\begin{equation}\label{tmat}
 {\cal T}_{\bf kk'}= \frac{F_\beta({\bf k})W_\beta(\omega)F_\beta({\bf k'})}
{1-W_\beta(\omega)G^0_{\beta\beta}(\omega)} .
\end{equation}
with the structure factor $F_\beta({\bf k})=\langle {\bf k}d|\beta\rangle$. 
Due to the predominantly $d$-character of the hole band, it is convenient to treat the problem
in the square lattice with a folded Brillouin zone (see Fig. 1 in the main text). Then the structure
factor is $F_\beta({\bf k})\approx 2(\cos k_x/2 - \cos k_y/2)$.

Next we estimate the contribution of the dangling bonds in the band
$\varepsilon_{{\bf k}p}$ neglected in the above calculations. This contribution would result
in a modification of the local Green's  function $G^0_{\alpha\beta}(\omega)$ in the effective
potential (\ref{spot}). Instead of the form $\langle\alpha| G_{\bf kk}|\alpha\rangle$ used 
in the right hand side of Eq.\ (\ref{tmat}), one should project the Green's  function on the subset
$\langle \beta|\ldots |\beta \rangle$ with the Slater-Koster defect in this band, namely, change
\begin{equation}\label{cor}
\langle\alpha| G_{\bf kk}|\alpha\rangle \to \langle \alpha| G_{\bf kk}\left(1+\frac{W_2
G_{\bf kk}}{1-W_2G^0_{\alpha\alpha}}\right)|\alpha\rangle
\end{equation}
(see Fig.\ 1 in the main text for the definition of $W_2$).

The first term in the right hand side of Eq.\ (\ref{cor}) is the Green's  function projected onto
the local $p$-orbitals $|\alpha\rangle$. Corrections due to the contribution of dangling bond states
in the $p$-band would be important only for those energies where 
$1-W_2G^0_{\alpha\alpha}(\omega) \sim 0$. 
However, it is seen from Eqs.\ (\ref{spot})  and (\ref{cor}) and from the shape of the 
DOS in Fig.\ 2  that the function $G^0_{\alpha\alpha}(\omega)$ is smooth in 
the region of overlap with the partial $d$ component of the DOS, and the singularities 
in the Green's  function reflecting the 2D van Hove singularities in the DOS
are located around the top of the $p$ band in the region of unoccupied states. In other words,
$W_2G^0_{\alpha\alpha}(\omega)\ll 1$ around the Fermi level, and the $p$ wave in Eq.\ (\ref{sec})
is represented by the unperturbed orbital $|\alpha\rangle$. 

\vspace{0.4cm}

\centerline{\bf Search for magnetic solutions}

\vspace{0.3cm} 

In order to find a magnetic solution, the Coulomb interaction is included in the scattering potential,
$W_\beta \to W_{\beta\sigma}= W_\beta+ U\bar n_{\beta,-\sigma}$.  
The average $\bar n^{}_{\beta\sigma}$ is given by 
\begin{eqnarray}
\bar n_{\beta\sigma} &=& \frac{1}{i\pi N}\sum_{{\bf k}_1{\bf k}_2}{\rm Im}\int^{\varepsilon_F}
F_\beta({\bf k}_1)F_\beta({\bf k}_2)
G_{{\bf k}_1{\bf k}_2,d\sigma}(\omega)d\omega 
  = \frac{1}{i\pi} \int^{\varepsilon_F}
G_{\beta\beta,\sigma}(\omega)d\omega. \ .
\end{eqnarray}
Its defect related part reads \cite{Midel67}
\begin{equation}
\bar n'_{\beta\sigma} = \frac{1}{i\pi U_{\beta\sigma}}\int^{\varepsilon_F} {\rm Im}
\left[\frac{1}{1- W_{\beta\sigma}G^0_{\beta\beta,\sigma}(\omega)} \right]d\omega.
\end{equation}
It is known \cite{Midel67,Wolff61}, that a magnetic solution 
$n_{\beta\uparrow}\neq n_{\beta\downarrow}$ exists
provided the repulsive potential exceeds some critical value,  
$W_{\beta c} > 1/ {\cal J}_c$ given by
\begin{equation}\label{instab}
{\cal J}_c =  \frac{1}{i\pi} \int^{\varepsilon_F} {\rm Im}\left\{
\frac{[-G^0_{\beta\beta}(\omega)]^2}{[1- W_{\beta c}G^0_{\beta\beta}(\omega)]^2} \right\}d\omega.
\end{equation}
Such a magnetic solution is expected to be 
realized due to the logarithmic singularity of $G^0_{\beta\beta}(\omega)$
in the very vicinity of $\varepsilon_F$. shown in Eq.\ (5) in the main 
text.  Due to this singularity in the denominator of the integrand in the r.h.s.\ of
Eq.\ (\ref{instab}) the factor ${\cal J}_c$ is strongly enhanced. This enhancement favors magnetic 
solutions.

The electron-band may be easily included in this calculation scheme, and the corresponding
secular matrix may be constructed in the same way as it was done above for the hole-band.
In order to describe the states in the
electron pockets (as well as the interband scattering), one should add at least one more orbital, 
namely the $d_{xy}$ one, to this minimal model \cite{YYO,Graser1,Graser2}.
However, the real part of the local Green's  function $G^0(\omega)<0$ near 
the bottom of the conduction band is negative \cite{Eco}, so the intraband 
scattering in the electron pocket is expected to be weak. Thus, we conclude 
that a $\rm V_{As}$ defect influences mainly the states 
in the hole pocket, leaving the electron pockets practically the same as
in perfect samples.\\ 

\centerline{\bf Few remarks on As-vacancies and spin-density wave states}

\vspace{0.3cm} 

Finally, we would like to mention an additional
challenging  
problem closely related to that of As-vacancies in a paramagnetic 
enviroment considered above: namely, 
the perturbational effect of As-vacancies on 
a surrounding spin density wave (SDW) to be investigated in more detail elsewhere.
From a general point of view it is however already
clear that sizable effects of common interest can be expected.
Indeed, since the competing spin-stripe SDW phase 
may be schematically represented
by a frustrated 2D $J$$_1$-$J$$_2$ or closely related spin model 
Hamiltonians
with an essential {\it antiferromagnetic }
next nearest neighbor
exchange coupling $J_2$
microscopically mediated 
by the superexchange involving the As-4$p$ states, the presence of an As-vacancy
will  locally eliminate this antiferromagnetic 
coupling $J_2$ even in the case of a nonmagnetic bound or resonance state
discussed above. Moreover, in case that the SDW (or its by doping weakened 
corresponding SDW-magnetic state) will not prevent the ferromagnetic 
polarization effect for the four sourrounding
Fe sites, the presence of such a local ferromagnetic "mini"-cluster as an extended
magnetic defect will obviously cause an additional weakening or destruction 
of the SDW or its fluctuations probably present
even in the "non-magnetic" superconducting state mentioned above.
This effect might explain the observed slight $T_c$-enhancement of about 2 to 3 K 
after creating
As-vacancies within an optimal doped system
(see Ref.\ 4 of the main text). Within a broader context magnetic
defects under control and As-vacancies in particular should provide a new 
tool to probe various
SDW states and this way give more insight into the complex
interplay of various competing ground states of Fe pnictides in general.


\begin{thebibliography}{99}
\bibitem{foot2} In the two-orbital model the exchange $J$ should also taken into account.
However, since $U>J$, it is enough to consider
the Coulomb intraband scattering in order to detect a magnetic solution.
The interband exchange gives only small
corrections
to this
result.
\bibitem{foot3} In our minimal model the doping effect is compensated by the levels split
off the top of hole band, but the third orbital $p_z$ may become the source of $el$-doping
and/or even contribute to the large local moment of
3.2$ \mu$$_{\rm B}$ per $V_{\rm As}$ according
to Ref.\ \onlinecite{Grine11}.
\end{thebibliography}
\end{document}